# A GN/EGN-MODEL REAL-TIME CLOSED-FORM FORMULA TESTED OVER 7,000 VIRTUAL LINKS

*M. Ranjbar Zefreh[1], A. Carena[1], F. Forghieri[2], S. Piciaccia[2], P. Poggiolini[1]*

[1]OptCom, Dipartimento di Elettronica e Telecomunicazioni, Politecnico di Torino, 10129, Torino, Italy
[2] Cisco Photonics Italy srl, via Santa Maria Molgora 48/C, 20871 Vimercate (MB), Italy
*pierluigi.poggiolini@polito.it*

**Abstract:** We derived a fully-closed-form GN-model formula and tested its accuracy of over 7,000 highly randomized C-band system scenarios. By further applying a correction that leverages the large system test-set, we were able to substantially improve it and obtain EGN-model accuracy with real-time-compatible computation effort.

## 1   Introduction

Physical-layer-aware control and optimization of ultra-high-capacity optical networks is becoming an increasingly important aspect of networking, as throughput demand and loads increase. A necessary pre-requisite to achieve it, is the availability of accurate analytical modeling of fiber non-linear effects (or NLI, Non-Linear-Interference). Several NLI models have been proposed, among which time-domain [1][2], GN [3], EGN [2][4][5], as well as [6]-[9], and others, including precursors of the above (see refs. in [10]). These models however either contain integrals that make them unsuitable for real-time use, or otherwise assume too idealized system set-ups. The challenge is to derive *approximate closed-form formulas* that both preserve accuracy and are general enough to model highly diverse actual deployed systems.

In the GN/EGN model class a rather general closed-form formula (CFF, Eqs. (41)-(43) in [3]) was available, which approximates the *incoherent* GN-model. We upgraded it to include, among other things, dispersion slope ($\beta_3$), following the approach shown in [11][12][13], and a term that models NLI self-coherence. This turns such CFF into a rather accurate and general *coherent GN-model* approximation.

We then tested the CFF over a 7,000 highly randomized C-band system test-set, encompassing: fully loaded and sparsely loaded combs; several PM-QAM and PM-Gaussian formats; three randomly mixed fiber types with randomly chosen span lengths. Testing was carried out by comparing the CFF with the full-fledged, numerically integrated EGN-model.

We then leveraged the large system test-set to embed into the CFF suitable best-fit correction factors to improve its accuracy. A substantial improvement in performance prediction accuracy was obtained this way, with the CFF essentially reproducing now the *EGN-model,* with minimal error. We then draw conclusions, based on the test results and on an assessment of the real-time computation performance of the CFF.

Preliminary results along similar research lines were presented in [14]. There, only PM-QAM formats were addressed, the coherence term was not present and the test-set encompassed far fewer (1,700) systems. Only 400 had frequency-dependent dispersion. General accuracy was substantially less than achieved here. This is ongoing research and the final target is that of extensive testing of the CFF with frequency-dependent loss as well as Interchannel Stimulated Raman Scattering (ISRS) over the (C+L) band, which are all elements supported by the models developed in [11][12][13].

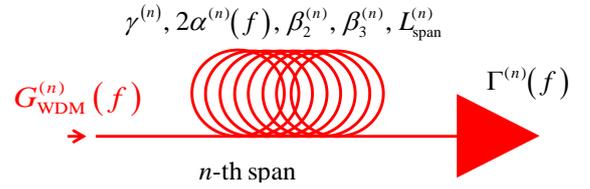

**Fig. 1:** Generic $n$-th span structure and related

## 2.   The closed-form model formula

The general structure of the considered link spans is shown in Fig.1. The superscript $(n)$ indicates that a quantity is related to the $n$-th span. Each span is composed of a single fiber type, with the customary parameters as indicated in Fig.1. At the end of the span there may be any combination of amplifiers, filters, and VOAs, whose aggregate transfer function is $\Gamma^{(n)}(f)$.

The WDM comb $G_{\mathrm{WDM}}^{(n)}(f)$ is fully arbitrary (Fig. 2) and can change span by span. Only the channel under test (CUT), which can be any of the channels, cannot change. Note that all quantities identified by a capital $G$ are power spectral densities

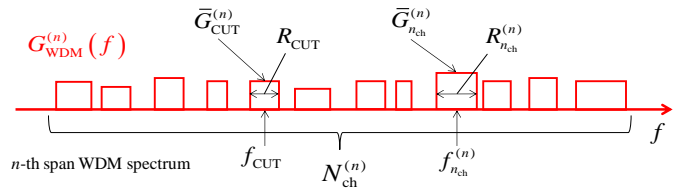

(PSD).

**Fig. 2:** generic $n$-th span WDM comb channel spectrum

Eqs. (1)-(5) make up the NLI-model CFF. The quantity $G_{\mathrm{NLI}}^{\mathrm{Rx}}(f_{\mathrm{CUT}})$ is the PSD of NLI at the frequency of the CUT. Assuming incoherent NLI accumulation, $G_{\mathrm{NLI}}^{\mathrm{Rx}}(f_{\mathrm{CUT}})$ is the sum of the PSD of NLI generated in each single span, $G_{\mathrm{NLI}}^{(n)}(f_{\mathrm{CUT}})$, as shown in Eq. (1). The sum is carried out after linear propagation of each of the $G_{\mathrm{NLI}}^{(n)}(f_{\mathrm{CUT}})$ terms till the end of the

$$G_{\text{NLI}}^{\text{Rx}}(f_{\text{CUT}}) = \sum_{n=1}^{N_{\text{span}}} \left( G_{\text{NLI}}^{(n)}(f_{\text{CUT}}) \prod_{k=n+1}^{N_{\text{span}}} \Gamma^{(k)}(f_{\text{CUT}}) \cdot e^{-2\alpha^{(k)}(f_{\text{CUT}}) \cdot L_{\text{span}}^{(k)}} \right) \quad (1)$$

$$G_{\text{NLI}}^{(n)}(f_{\text{CUT}}) = \frac{16}{27}(\gamma^{(n)})^2 \Gamma^{(n)}(f_{\text{CUT}}) \cdot e^{-2\alpha^{(n)}(f_{\text{CUT}}) \cdot L_{\text{span}}^{(n)}} \cdot \bar{G}_{\text{CUT}}^{(n)} \cdot \left( \rho_{\text{CUT}}^{(n)} \cdot \left[\bar{G}_{\text{CUT}}^{(n)}\right]^2 I_{\text{CUT}}^{(n)} + \sum_{n_{\text{ch}}=1,\, n_{\text{ch}} \neq n_{\text{CUT}}^{(n)}}^{N_{\text{ch}}^{(n)}} 2\rho_{n_{\text{ch}}}^{(n)} \cdot \left[\bar{G}_{n_{\text{ch}}}^{(n)}\right]^2 I_{n_{\text{ch}}}^{(n)} \right) \quad (2)$$

$$I_{\text{CUT}}^{(n)} = \frac{\operatorname{asinh}\left(\frac{\pi^2}{2}\left|\frac{\bar{\beta}_{2,\text{CUT}}^{(n)}}{2\alpha^{(n)}(f_{\text{CUT}})}\right| R_{\text{CUT}}^2\right) + \frac{2\,\text{Si}\left(\pi^2 \bar{\beta}_{2,\text{CUT}}^{(n)} L_{\text{span}}^{(n)} B_{\text{CUT}}^2\right)}{\text{Si}_{\max} 2\alpha_n(f_{\text{CUT}})L_{\text{span}}^{(n)}}\left[HN(N_{\text{span}}-1) + \frac{1-N_{\text{span}}}{N_{\text{span}}}\right]}{2\pi\left|\bar{\beta}_{2,\text{CUT}}^{(n)}\right| \cdot 2\alpha^{(n)}(f_{\text{CUT}})} \quad (3)$$

$$I_{n_{\text{ch}}}^{(n)} = \frac{\operatorname{asinh}\left(\pi^2 \left|\frac{\bar{\beta}_{2,n_{\text{ch}}}^{(n)}}{2\alpha^{(n)}\left(f_{n_{\text{ch}}}^{(n)}\right)}\right|\left[f_{n_{\text{ch}}}^{(n)} - f_{\text{CUT}} + \frac{R_{n_{\text{ch}}}^{(n)}}{2}\right] R_{\text{CUT}}\right) - \operatorname{asinh}\left(\pi^2 \left|\frac{\bar{\beta}_{2,n_{\text{ch}}}^{(n)}}{2\alpha^{(n)}\left(f_{n_{\text{ch}}}^{(n)}\right)}\right|\left[f_{n_{\text{ch}}}^{(n)} - f_{\text{CUT}} - \frac{R_{n_{\text{ch}}}^{(n)}}{2}\right] R_{\text{CUT}}\right)}{4\pi\left|\bar{\beta}_{2,n_{\text{ch}}}^{(n)}\right| \cdot 2\alpha^{(n)}\left(f_{n_{\text{ch}}}^{(n)}\right)} \quad (4)$$

$$\bar{\beta}_{2,\text{CUT}}^{(n)} = \beta_2^{(n)} + \pi\beta_3^{(n)}\left[2f_{\text{CUT}} - 2f_c^{(n)}\right] \quad , \quad \bar{\beta}_{2,n_{\text{ch}}}^{(n)} = \beta_2^{(n)} + \pi\beta_3^{(n)}\left[f_{n_{\text{ch}}}^{(n)} + f_{\text{CUT}} - 2f_c^{(n)}\right] \quad (5)$$

link, accounted for by the product symbol in Eq. (1). Eq. (2) provides the $G_{\text{NLI}}^{(n)}(f_{\text{CUT}})$, which consist of a self-channel interference (SCI) and a cross-channel interference (XCI) contribution to NLI. These are expressed as integrals in the GN-model, whose approximate closed-form solutions are $I_{\text{CUT}}^{(n)}$ Eq. (3) and $I_{n_{\text{ch}}}^{(n)}$ Eq. (4), respectively. In these equations, $n_{\text{ch}}$ is the channel number, ranging from 1 to $N_{\text{ch}}^{(n)}$ in the $n$-th span. Frequency-dependent dispersion can be modeled through the $\beta_3$ coefficient. The inclusion of $\beta_3$ in the GN-model equations was shown in [10] Eq. (C2). We started off from that result. However, to close the integrals, we had to assume that dispersion is different at each channel center frequency, but constant over the bandwidth of each channel. This approach is the same as used in [11][12][13]. A similar assumption was made for the loss coefficient $\alpha(f)$. If Interchannel Stimulated Raman Scattering (ISRS) is included as well, more complex but still closed-form expressions can be found [11][12][13] (not addressed here). Testing with ISRS is work in progress.

The term in Eq. (3) that is circled in gray is the *NLI coherence term*. It effectively turns Eqs. (1)-(5) from an incoherent GN-model into a *coherent* GN-model CFF approximation [15]. The factors $\rho_{\text{CUT}}^{(n)}$ and $\rho_{n_{\text{ch}}}^{(n)}$ in Eq. (2) are correction terms that further transform Eqs. (1)-(5) into an *EGN-model* CFF approximation, as shown later.

## 2. Test-Set Generation and Model Testing

The 7,000 systems test-set was generated as follows. In about 5,400 systems the C-band was fully populated. In the remainder, at launch we turned off anyone of the WDM channels with probability 1/2, so that the average load was 50% and actual loads were widespread. In all 7,000 systems the symbol rate of each channel (including the CUT) was randomly chosen among 32, 64, 96 and 128 GBaud. Also, the format was randomly chosen among 16, 32, 64, 128, 256 PM-QAM and PM-Gaussian. The roll-off factor was uniformly distributed between 0.05 and 0.25. Channel spacing was random, with lower limit of no-interference and upper limit 43.5, 87.5, 131.25, 175 GHz, for 32, 64, 96 and 128 GBaud, respectively. 1/3 of the systems had as CUT the center channel, 1/3 the lowest and 1/3 the highest frequency channel. For PM-QAM CUTs, the target operating OSNR was chosen so that in AWGN it would results in a normalized GMI of 0.87. For PM-Gaussian CUTs the target OSNRs were randomly chosen between those yielding the same normalized GMI assumed for PM-16QAM and that of PM-256QAM. The WDM combs extended over 5 THz in the range $f_c \pm 2.5$ THz, with $f_c$= 193.415 THz.

The fiber type was randomly chosen for *each span*, among three fiber types: SMF, NZDSF1 and NZDSF2, whose parameters were, respectively: $\alpha$, 0.21, 0.22, 0.22 dB/km; $\beta_2$, -21.3, -4.85, -2.59 ps$^2$/km; $\beta_3$, 0.1452, 0.1463, 0.1206 ps$^3$/km; $\gamma$, 1.3, 1.35, 1.77 1/(W km). Each span length was generated according to a uniform distribution between 80 and 120 km. The EDFAs noise figure was randomized between 5 and 6 dB. The nominal launch power of each channel was optimized *at each span* according to the LOGO strategy [3] Eq. (82), but then each channel launch power (except for the CUT) was randomly altered between 70% to 130% of the optimum power, to mimic real-system power imbalances.

As reference we used the full-fledged, numerically integrated EGN-model, in the version [4], which we set as *benchmark*. This model has repeatedly proved very accurate in several extensive validations, such as [16][17]. The test procedure was as follows. For each test-set system, first the max-reach was found using the benchmark EGN-model. Then, at max-reach, the quantity $\text{OSNR}_{\text{NL}} = P_{\text{ch}}/(P_{\text{ASE}} + P_{\text{NLI}})$ was estimated, both with the benchmark EGN-model, yielding $\text{OSNR}_{\text{NL}}^{\text{EGN}}$, and with Eqs.(1)-(5), giving $\text{OSNR}_{\text{NL}}^{\text{CFF}}$. Then the error was assessed as:

$$ERR = \left(\text{OSNR}_{\text{NL}}^{\text{CFF}}/\text{OSNR}_{\text{NL}}^{\text{EGN}}\right)_{\text{dB}}$$

The plot of *ERR* over all 7,000 systems is shown in Fig. 3. The red histograms were found with the correction factors $\rho_{\text{CUT}}^{(n)}$ and $\rho_{n_{\text{ch}}}^{(n)}$ set to 1 (no correction). Despite the extreme diversity of the systems considered, and the drastic approximations used to obtain Eqs. (1)-(5), the standard deviation of *ERR* (see Fig. 3 insets) is only about 0.2 dB, a remarkably small value. There is however a relatively large mean value shift, of about 0.65 dB, attributable in part to the GN-model (of which Eqs.(1)-(5) are an approximation) tendency to overestimate NLI.

To improve accuracy, we then used the correction factors defined as shown in Eq. (6). They involve the following physical quantities for the CUT and for each channel $n_{\text{ch}}$: the roll-off factors $r$; the EGN-model format-dependence constant $\Phi$ (whose values are listed in [4][10]); the effective

$$\rho_{\text{CUT}}^{(n)} = (1 + a_9 \cdot r_{\text{CUT}}^{a_{10}})(a_{11} + a_{12} \cdot \Phi_{\text{CUT}}^{a_{13}} + a_{21} \cdot \Phi_{\text{CUT}}^{a_{22}} \cdot (1 + a_{14} R_{\text{CUT}}^{a_{15}} + a_{16}(|\bar{\beta}_{2,\text{acc}}(n, n_{\text{CUT}})| + a_{17})^{a_{18}}))$$
$$\rho_{n_{\text{ch}}}^{(n)} = (1 + a_{23} \cdot n_{\text{ch}}^{a_{24}})(1 + a_1 \cdot r_{\text{CUT}}^{a_2})(a_3 + a_4 \cdot \Phi_{n_{\text{ch}}}^{a_5} + a_{19} \cdot \Phi_{n_{\text{ch}}}^{a_{20}}(1 + a_6(|\bar{\beta}_{2,\text{acc}}(n, n_{\text{ch}})| + a_7)^{a_8}))$$
$$\bar{\beta}_{2,\text{acc}}(n, n_{\text{ch}}) = \sum_{k=1}^{n-1} \bar{\beta}_{2,n_{\text{ch}}}^{(k)} \cdot L_{\text{span}}^{(k)} \qquad (6)$$

accumulated dispersion $\bar{\beta}_{2,\text{acc}}(n, n_{\text{ch}})$ at span $n$, for channel $n_{\text{ch}}$ or for the CUT (see also Eq.(5)). There are then the free parameters $a_1$ to $a_{24}$. For their best-fitting, we used a standard MSE minimization algorithm on the quantity *ERR*, looking at only 2,500 out of the 7,000 test-set systems to avoid possible risk of overfitting. The resulting values for $a_1$ to $a_{24}$ were: [-1.6139, 2.6360, 0.9653, -1.36211, 0.84213, -1.02231, 5.38270, 3.77720e-3, -1.08013, 1.91066, 0.88153, -2.66093, 1.4050, -1.11174, 7.3518e-3, 2.60510e8, 2.24475e3, -3.02058, -19.4215, 0.847, -28.04338, 1.52887, -1.42818, 1.91285].

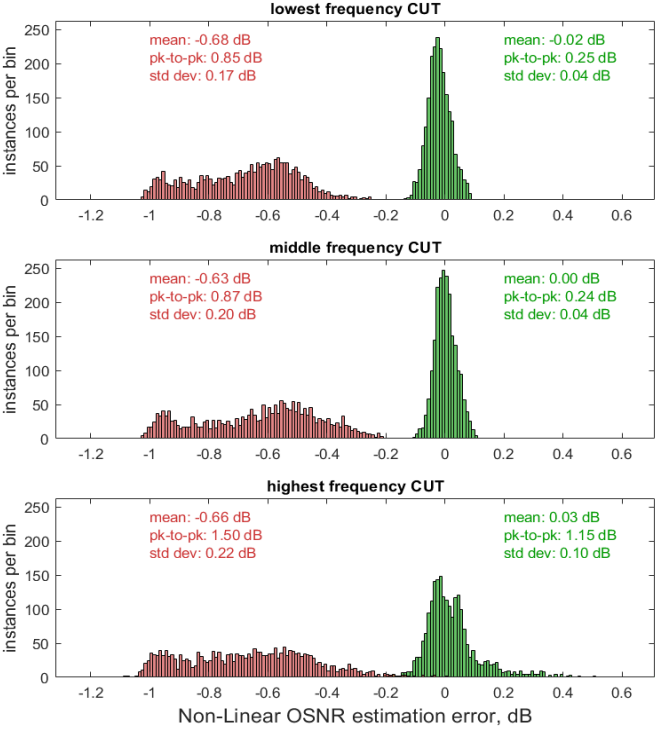

**Fig. 3:** Histogram of the non-linear OSNR estimation error: $ERR = (\text{OSNR}_{\text{NL}}^{\text{CFF}}/\text{OSNR}_{\text{NL}}^{\text{EGN}})_{\text{dB}}$, where $\text{OSNR}_{\text{NL}}^{\text{EGN}}$ is found using the benchmark EGN-model and $\text{OSNR}_{\text{NL}}^{\text{CFF}}$ is found with the closed-form formula Eqs. (1)-(5), at system maximum reach. Total of 7,000 systems over the three plots. Green histograms: with EGN-correction; red histograms: without EGN-correction.

As a result of introducing the corrections, we obtained the green histograms for the error, shown in Fig. 3. The error of the CFF Eqs. (1)-(5) vs. the EGN-model benchmark, on the lowest and center frequency of the comb, has now a negligible mean, a negligible standard deviation of 0.04 dB and a peak-to-peak error across all systems of only 0.25 dB.

However, when looking at the high-frequency channel, although the mean is still essentially zero and the standard deviations is only 0.1 dB, a peak-to-peak error of 1.15 dB is recorded, which shows that there are now *outliers*. Note that some are so infrequent to be invisible in figure. The reason for the presence of these outliers is shown in Fig. 4, where the WDM comb of one the outlier cases is shown. The CUT was the high-frequency-channel, transmitting 128 PM-QAM. The max-reach was 2 spans and both were NZDSF2. Therefore, transmission occurs entirely at $D= 0.58$ ps/(nm km). At such low values of dispersion some of the CFF approximations break down and the CFF error vs. the benchmark EGN-model goes up. We performed further tests and found that it must be $D>1$ ps/(nm km) to avoid outlier cases.

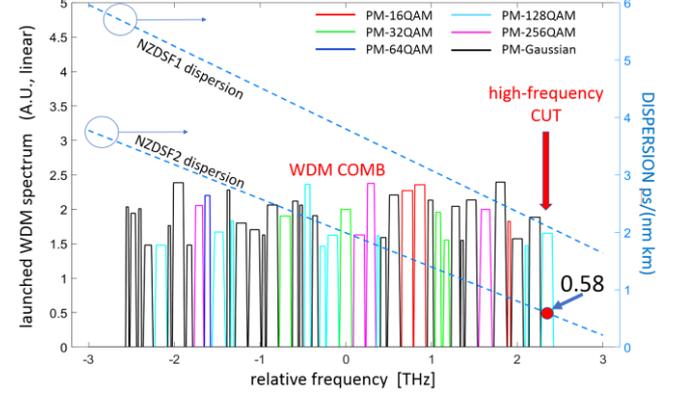

**Fig. 4:** Solid lines: one of the launched WDM test combs. Full C-band (5 THz). Different channel colors correspond to different formats (see legend). Dashed curves: dispersion plots for fibers NZDSF1 and NZDSF2.

As a last important test, we estimated the computation time of the CFF. It took on average about 15 ms to calculate the NL-OSNR for *all* WDM channels of a system, using a laptop and interpreted Matlab[TM] code. This is several orders of magnitude faster than using numerical integration of the EGN or even GN-model and adequate for real-time use.

## 3. Conclusion

We tested the accuracy of a closed-form approximate GN-model formula over 7,000 highly-randomized C-band system set-ups (5400 fully-loaded and 1,600 partially-loaded), that used a variety of PM-QAM and PM-Gaussian formats, different spacings, different symbol rates and 3 very different fiber types. To the best of our knowledge, this is the first time such an extensive study has been performed.

We then greatly improved the accuracy of the model by leveraging the large test-set available to obtain correction factors that weigh appropriately several physically relevant system parameters. Away from pathological near-zero-dispersion situations, the formula showed very good accuracy in reproducing the results of the full-fledged, numerically-integrated EGN-model, at a comparatively negligible computational effort.

We therefore believe the approximate closed-form formula proposed here could potentially provide an effective and accurate tool to support real-time physical-layer-aware management and control of optical networks.

Further model upgrades to improve accuracy as well as testing over C+L band (with ISRS), are in progress

This work was supported by Cisco Systems through an SRA contract and by the PhotoNext Center of Politecnico di Torino.

# 4 References

[1] A. Mecozzi and R.-J. Essiambre, 'Nonlinear Shannon limit in pseudolinear coherent systems,' Journal of Lightwave Technology, vol. 30, no. 12, pp. 2011–2024, June 15th 2012.

[2] R. Dar, M. Feder, A. Mecozzi, and M. Shtaif, 'Properties of nonlinear noise in long, dispersion-uncompensated fiber links,' Optics Express, vol. 21, no. 22, pp. 25685–25699, Nov. 2013.

[3] P. Poggiolini, G. Bosco, A. Carena, V. Curri, Y. Jiang, F. Forghieri, 'The GN model of fiber non-linear propagation and its applications,' Journal of Lightwave Technology, vol. 32, no. 4, pp. 694–721, Feb. 2014.

[4] A. Carena, G. Bosco, V. Curri, Y. Jiang, P. Poggiolini and F. Forghieri, 'EGN model of non-linear fiber propagation,' Optics Express, vol. 22, no. 13, pp. 16335–16362, June 2014.

[5] P. Serena, A. Bononi, 'A Time-Domain Extended Gaussian Noise Model,' Journal of Lightwave Technology, vol. 33, no. 7, pp. 1459–1472, Apr. 2015.

[6] A. Bononi, P. Serena, N. Rossi, E. Grellier, F. Vacondio, 'Modeling nonlinearity in coherent transmissions with dominant intrachannel-four-wave-mixing,' Optics Express, vol. 20, pp. 7777-7791, 26 March 2012.

[7] P. Johannisson, M. Karlsson, 'Perturbation analysis of nonlinear propagation in a strongly dispersive optical communication system,' Journal of Lightwave Technology vol. 31, no. 8, pp. 1273-1282, Apr. 15, 2013.

[8] M. Secondini and E. Forestieri, 'Analytical fiber-optic channel model in the presence of cross-phase modulations,' IEEE Photonics Technology Letters, vol. 24, no. 22, pp. 2016–2019, Nov. 15th, 2012.

[9] R. Dar, M. Feder, A. Mecozzi, M. Shtaif, 'Pulse collision picture of inter-channel nonlinear interference noise in fiber-optic communications,' Journal of Lightwave Technology, vol. 34, no. 2, pp. 593–607, Jan. 2016.

[10] P. Poggiolini, Y. Jiang, A. Carena, F. Forghieri 'Analytical Modeling of the Impact of Fiber Non-Linear Propagation on Coherent Systems and Networks,' Chapter 7 in: *Enabling Technologies for High Spectral-efficiency Coherent Optical Communication Networks*, p. 247-310, Wiley, ISBN: 978-111907828-9, doi: 10.1002/9781119078289.

[11] D. Semrau, R. I. Killey, P. Bayvel, 'A Closed-Form Approximation of the Gaussian Noise Model in the Presence of Inter-Channel Stimulated Raman Scattering,' www.arXiv.org, paper arXiv:1808.07940, Aug. 23rd 2018.

[12] P. Poggiolini 'A generalized GN-model closed-form formula,' www.arXiv.org, paper arXiv:1810.06545v2, Sept. 24th 2018.

[13] D. Semrau, R. I. Killey, P. Bayvel, 'A Closed-Form Approximation of the Gaussian Noise Model in the Presence of Inter-Channel Stimulated Raman Scattering,' Journal of Lightwave Technology, vol. 37, no. 9, pp. 1924–1936, May 1st, 2019.

[14] P. Poggiolini, M. Ranjbar Zefreh, G. Bosco, F. Forghieri, S. Piciaccia, 'Accurate Non-Linearity Fully-Closed-Form Formula based on the GN/EGN Model and Large-Data-Set Fitting,' in Proc. of OFC 2019, paper M1I.4, San Diego (CA), Mar. 2019.

[15] P. Poggiolini 'A closed-form GN-model Non-Linear Interference coherence term,' www.arXiv.org, April 2019.

[16] P. Poggiolini, Y. Jiang, 'Recent Advances in the Modeling of the Impact of Nonlinear Fiber Propagation Effects on Uncompensated Coherent Transmission Systems,' tutorial review, J. of Lightwave Technol., vol. 35, no. 3, pp. 458-480, Feb. 2017.

[17] P. Poggiolini et al, 'Non-Linearity Modeling for Gaussian-Constellation Systems at Ultra-High Symbol Rates' ECOC 2018, paper Tu4G.3.